\documentclass[preprint,showpacs,preprintnumbers,amsmath,amssymb]{revtex4}

\usepackage{graphicx}
\usepackage{dcolumn}
\usepackage{bm}

\begin{document}
 
 \author{M.~E.~Povarnitsyn$^{1,2}$, T.~E.~Itina$^1$, M.~Sentis$^1$,  K.~V.~Khishchenko$^2$, and P.~R.~Levashov$^2$}
 \affiliation{$^{1}$Laboratory of Lasers, Plasmas  and Photonic Processing (LP3, UMR 6182 CNRS), 163 avenue de Luminy, Case 917, 13288, Marseille, France \\ $^2$Institute for High Energy Densities RAS, Izhorskaya 13/19, Moscow, 125412, Russia}

\title{Material decomposition mechanisms in femtosecond laser interactions with metals}
\date{\today}

\begin{abstract}
A numerical hydrodynamic study of femtosecond laser ablation is presented.  A detailed analysis of material decomposition is performed using a thermodynamically complete equation of state with separate stable and metastable phase states and phase boundaries.  The lifetime of the metastable liquid state is estimated based on the classical theory of homogeneous nucleation.  In addition, mechanical fragmentation of the target material is controlled based on available criteria.  As a result, several ablation mechanisms are observed.  A major fraction of the ablated material, however, is found to originate from the metastable liquid region, which is decomposed either \textit{thermally} in the vicinity of the critical point into a liquid--gas mixture, or \textit{mechanically} at high strain rate and negative pressure into liquid droplets and chunks.  The calculation results explain available experimental findings. 
\end{abstract}


\maketitle
\section{INTRODUCTION}
Femtosecond laser interactions have attracted a particular interest of researchers during the last decade.  It is not surprising since these new ultra-short laser pulses provide a unique opportunity for the fast development of numerous very promising and exciting applications, such as laser machining, micro and nano-structuring, laser-induced plasma spectroscopy, nanoparticle synthesis in vacuum, in gas or in a liquid solution, medical imaging, laser surgery, etc. \cite{R1,R2,R3}  Many experimental studies were performed demonstrating the unique possibilities of femtosecond laser systems. \cite{Nolte,Amoruso,Plech}  At the same time, several theoretical investigations were focused on the better understanding of physical processes involved in ultra-short laser interactions. \cite{Colombier05, Colombier06, Rethfeld,Anisimov1,Komashko,Ivanov, Anisimov2, Eidmann, Vidal, Glover, Zhigilei, Garrison, Schafer, Perez, Lorazo}  The classical approach typically used for these interactions is based on a two-temperature model. \cite{Anisimov2}  In addition, several hydrodynamic simulations were carried out to describe the target material motion. \cite{Anisimov1,Komashko, Eidmann, Vidal, Glover}  The formation of a thermal wave, pressure gain and shock wave propagation have been observed in these calculations.  Molecular dynamics (MD) simulations were furthermore performed that provided insights into the mechanisms, such as \textit{phase explosion}, \textit{fragmentation}, \textit{evaporation}, and \textit{mechanical spallation}. \cite{Zhigilei, Garrison, Schafer, Perez, Lorazo}

In spite of a significant effort aimed at the understanding of the ablation mechanisms, several issues stay unclear.  For example, theoretically predicted ablation depth was frequently underestimated with respect to the typical experimental values. This result is an indication that not all the ablation mechanisms were taken into account.  In addition, numerous terms used to describe these mechanisms are confusing or misused.  For instance, when \textit{explosive boiling} term is used, which reflects the overheated (metastable) liquid decomposition into a liquid--gas mixture, it is often associated with material heating.  The metastable liquid state, however, can be also achieved during fast material expansion and cooling.  The proper term for this process is unclear.  In this case, \textit{critical-point phase separation} was proposed in Ref.~\onlinecite{Vidal}, but only to describe material decomposition in the vicinity of the critical point.  In addition, up to now, it was difficult to handle this process correctly in the framework of hydrodynamic models, because most of the available equations of state (EOS) did not describe the metastable regions.  Moreover no kinetics of gas bubble formation or/and condensation was previously introduced to determine the lifetime of these states.  Besides, there is still a lack of information about the photomechanical failure of the material at extremely high strain rates (up to $10^9$~s$^{-1}$), which can be achieved. \cite{Colombier05} Nevertheless, even these high strain rates do not always satisfy the fragmentation criteria \cite{Grady, Kanel} because of their short duration (tens of picosecond). All these issues need more analysis. 
\section{MODEL}
To bring more light on the mechanisms of material decomposition in femtosecond laser ablation of metals, we performed a detailed hydrodynamic modeling. Remind, that compared to atomistic techniques, hydrodynamic models allow calculations for much larger systems and take much shorter computer time. Our numerical model is based on the solution of a system of Eulerian hydrodynamic equations by a high-order multi-material Godunov's method. \cite{Miller}  The equations were extended to the case of one-fluid two-temperature hydrodynamics with laser energy absorption source, electron heat conductivity and energy exchange between electrons and heavy particles (atoms, ions, nuclei). \cite{Povarnitsyn}  A Gaussian temporal profile is used to simulate the laser energy deposition. The heat conductivity of electrons is calculated according to model. \cite{Anisimov3} Energy exchange between electrons and heavy particles, the reflectivity coefficient $R$ and the optical penetration depth $\lambda_{opt}$ are derived by means of the wide-range frequency of electron-phonon/ion collisions. \cite{Eidmann}

For completeness of our model, we use a semiempirical multi-phase EOS for aluminum with separate description of subsystems of heavy particles and electrons. The specific Helmholtz free energy has a form $\mathcal{F}(\rho, T_i, T_e)=\mathcal{F}_i(\rho,T_i)+\mathcal{F}_e(\rho, T_e)$, composed of two parts, which describe the contributions of heavy particles and electrons, respectively. Here $\rho$ is the material density, $T_i$ and $T_e$ are temperatures of heavy particles and electrons. The first item $\mathcal{F}_i(\rho,T_i) =\mathcal{F}_c(\rho) +\mathcal{F}_a(\rho, T_i)$, in turn, consists of the electron-ion interaction term $\mathcal{F}_c$ (calculated at $T_i=T_e=0$~K) and the contribution of thermal motion of heavy particles $\mathcal{F}_a$. Analytical form of $\mathcal{F}_i$ has different expressions for the solid $\mathcal{F}_i^{(s)}$, as well as for both liquid and gas phases $\mathcal{F}_i^{(l)}$. \cite{Khishchenko} Using these thermodynamic functions, the solid, liquid, and gas phases equlibrium boundaries are determined from the equality conditions for the temperature $T_i$, pressure $P_i=\rho^2(\partial\mathcal{F}_i/\partial\rho)_{T_i}$, and Gibbs potential $G_i=\mathcal{F}_i+P_i/\rho$ of each phases pairs. \cite{LandauLifshits} The tables of thermodynamic parameters are calculated taking into consideration the information about phase transitions and metastable regions. \cite{Khishchenko1, Oreshkin} The free energy of electrons in metal $\mathcal{F}_e$ has a finite-temperature ideal Fermi-gas form. \cite{LandauLifshits} 

The model of our EOS contains about 40 adjustable constants so that the EOS meets the following requirements: (i) to describe experimental results on compression and expansion for a wide range of densities and temperatures including data on critical and triple points; (ii) to contain separate information about electron and ion/lattice subsystems; (iii) to represent changes of thermodynamic parameters during phase transitions. To account for kinetic processes, an estimation of the realistic lifetime of metastable liquid state is introduced. Using this lifetime-based information we switch between two different modifications of the EOS: (i) with metastable states; and (ii) without metastable states.
 
In our model, when the liquid branch of the binodal curve was crossed and the matter turned into metastable state, we include a particular treatment for each of the following two competitive effects: (i) for the thermal decomposition, a criterion of the metastable liquid lifetime is used based on the theory of homogeneous nucleation; \cite{Frenkel,Skripov,Tkachenko} (ii) for the mechanical fragmentation, a  failure criterion of Grady is applied. \cite{Grady}  

In the first case, we estimate the metastable liquid lifetime as $\tau_{nucl}=(CnV)^{-1}\textnormal{exp}(W/k_BT_i)$, where $C=10^{10}$~s$^{-1}$ is the kinetic coefficient, \cite{Tkachenko} $n$ is the concentration, $V$ is the volume under consideration (the volume of numerical cell in our case), $W=16\pi\sigma^3/3\Delta P^2$  is the work required on a gas bubble formation in a liquid phase, $k_B$ is the Boltzmann constant, $\Delta P$ is the difference between the saturated vapor pressure, known from EOS, and the pressure of substance.  The temperature dependence of the surface tension $\sigma$ is described in the form $\sigma=\sigma_0(1-T_i/T_c)^{1.25}$, see Ref.~\onlinecite{Surftens,Tkachenko}. Here $T_c=6595$~K is the temperature in critical point (CP) for Al, \cite{Khishchenko} $\sigma_0=860$~g/s$^2$ is the surface tension of Al at melting temperature. \cite{Surftens}  In this case, as soon as the lifetime $\tau_{nucl}$ in the volume $V$ is expired, the metastable one-phase state separates into two-phase stable liquid--gas mixture. This phase separation process is accompanied by abrupt changes of thermodynamic parameters such as pressure, temperature, sound speed, compressibility, heat capacity, etc. The EOS with metastable phase states is therefore no more relevant in this volume, so that we continue to calculate the thermodynamic properties by using the EOS without metastable states.

To account for the second effect, a fragmentation criterion \cite{Grady} is used for the liquid phase with the spall strength $P_{spal}=(6\rho^2 c^3\sigma\dot{\varepsilon})^{1/3}$ and the time to fracture $\tau_{spal}=(6\sigma/\rho\dot{\varepsilon}^2 c^3)^{1/3}$, where $\dot{\varepsilon}$ is the strain rate and $c$ is the sound speed.  When the pressure of matter drops below the negative value $-P_{spal}$ and duration of this event is longer than the time $\tau_{spal}$, a criterion of fragmentation is satisfied and voids and new free surfaces can appear in the substance. To describe this phenomenon numerically we let a ``broken'' substance to shrink back until the pressure comes to zero value. The difference between old and new volumes we compensate by introducing void fraction into a numerical cell.  Both thermal and mechanical criteria described above are used simultaneously and each of them can prevail in a given computational cell depending on the substance location on the phase diagram.
\section{RESULTS}
\begin{figure}
\includegraphics[width=1.0\columnwidth]{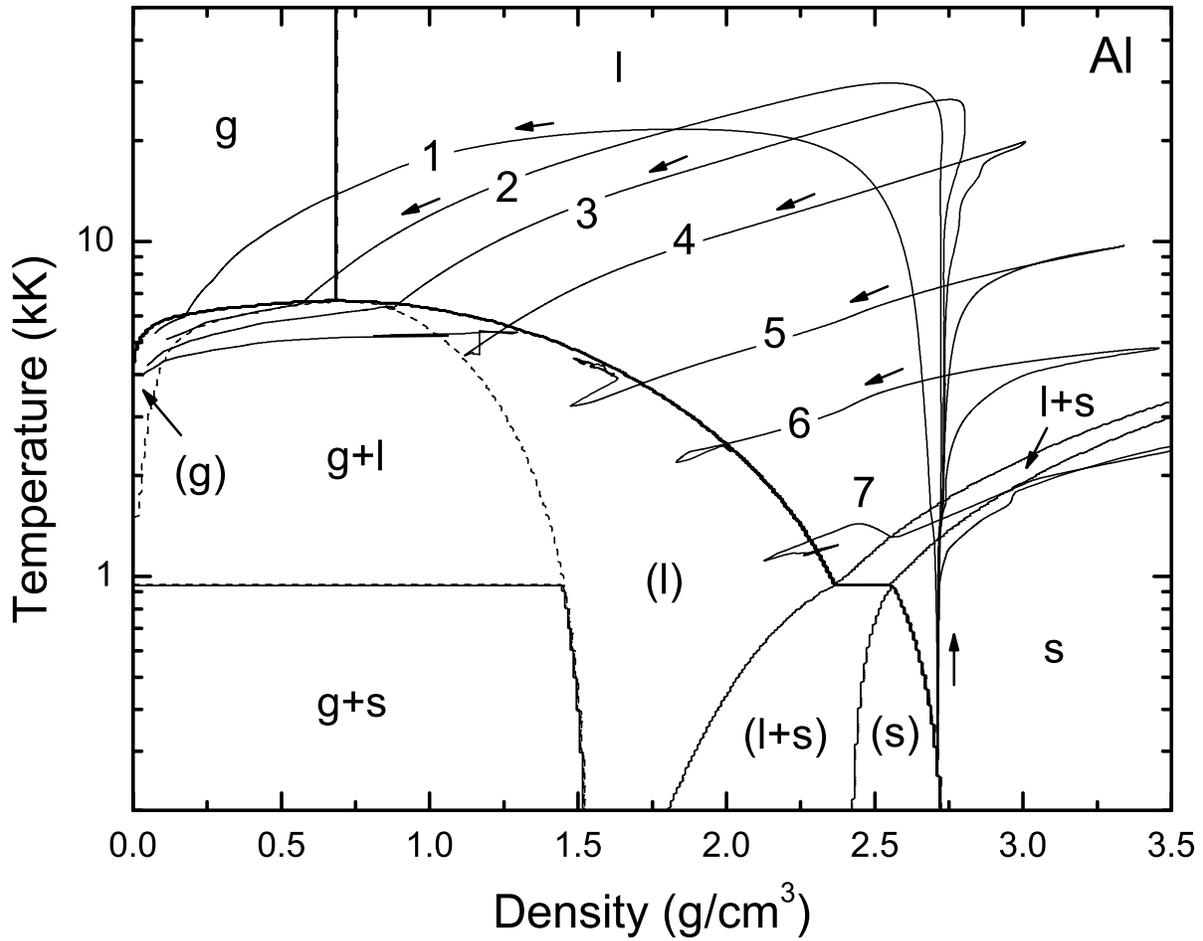}
\caption{\label{fig:1} Phase diagram of aluminum and time evolution of density and temperature of heavy particles for different layers of the target. Dashed curve: the spinodal; g: stable gas; l: stable liquid; s: stable solid; l+s: stable melting; g+l: liquid--gas mixture; g+s: solid--gas mixture; (g): metastable gas; (l): metastable liquid; (l+s): metastable melting; (s): metastable solid. The phase trajectories 1--7 correspond to depths of 5, 15, 20, 30, 50, 80, 130~nm from the initial target surface, respectively. Arrows along the trajectories show the flow of time. Here, laser pulse parameters are $\tau_L=100$~fs, $\lambda_L=800$~nm, and $F=5$~J/cm$^2$.}
\end{figure}

For the analysis of the target material evolution after the laser irradiation, it is convenient to use a phase diagram in the temperature-density plane $(T, \rho)$ given by the EOS (Fig.~\ref{fig:1}).  
The binodal curve corresponds to the liquid--vapor coexistence line (saturated vapor curve). The spinodal curve (dashed) gives a limit of thermodynamic stability of the liquid and gas phases. The regions between the binodal and the spinodal curves show the metastable states of matter: the superheated liquid (l) and the supercooled vapor (g) states.  

Initially, the aluminum target is in a solid state with $\rho=2.71$~g/cm$^3$, $T_i=T_e=300$~K, and is subjected to a laser pulse with fluence $F=5$~J/cm$^2$, pulse width $\tau_{L}=100$~fs and wavelength $\lambda_L=800$~nm.  The reflectivity coefficient and the optical penetration depth are calculeted~\cite{Eidmann} and their values are $R=0.92$ and $\lambda_{opt}=13$~nm, respectively. The conduction-band electrons absorb laser energy in a skin-layer (a very thin region of the order of the optical penetration depth $\lambda_{opt}$). Then, these hot electrons transmit absorbed energy into the bulk owing to heat conduction and warm up the lattice due to electron-phonon collisions. The time of electron-lattice relaxation (equilibration of temperatures of electron and heavy particle subsystems) is of the order of several picoseconds for aluminum and thus the lattice pressure has time to grow up before the heat energy of electrons dissipates into the bulk of the target. Formation of a narrow zone with a sharp pressure profile results in generation of intensive shock wave, which moves into the bulk, and rarefaction wave that causes a spread of the matter out of the target. The process of the shock wave propagation is clearly visible as the consecutive right-hand deviations of the 30, 50, 80, 130~nm-trajectories in Fig.~\ref{fig:1}. Process of compression of different target layers changes into expansion one (decreasing of the density along trajectories, Fig.~\ref{fig:1}).  

The maximal temperature of heavy particles in the vicinity of the target surface (depth $\le15$~nm) is high enough (up to $T_i\sim30$~kK), so that the phase trajectories from this layer go above the CP and the target material is directly transformed into the gas phase (atomization, 5 and 15~nm trajectories in Fig.~\ref{fig:1}). Then, these trajectories cross the gas branch of the binodal and penetrate into the supercooled vapor region, where condensation starts and  liquid--gas mixture forms.  Note, however, that this layer represents a very small fraction of the ablated material, so that the condensation degree is insignificant.  

Next layer in the target (depth from 20 to 30~nm) is first transformed into a metastable liquid state. This layer is heated to high temperatures (up to $T_i\sim25$~kK), so that the corresponding phase trajectories cross the liquid branch of binodal in the vicinity of the CP and enter the metastable liquid region.  Under these conditions, the target material is thermodynamically unstable. The lifetime of this state is estimated as described above. Thermodynamic instabilities are known to occur near the CP (particularly, at $0.9 T_c < T_i < T_c$) leading to a rapid decomposition (several picoseconds) of the matter into a liquid--gas mixture.
This process is similar to the \textit{phase explosion}, \cite{Zhigilei, Garrison} though occurs during simultaneous material expansion and cooling.  The other term used for this mechanism is \textit{critical-point phase separation} proposed in Ref.~\onlinecite{Vidal}.  

This thermal mechanism, however, concerns only quite a thin slice ($\sim10$~nm thickness), which trajectories enter the metastable liquid region near the critical point.  In fact, the melted layer is much thicker ($\sim200$~nm at $F=5$~J/cm$^2$).  The rest of the trajectories (5--7 in Fig.~\ref{fig:1}), which originate from the deeper lying melted layers of the target, also enter the metastable liquid region, but at temperatures much smaller than the critical temperature $T_c$.  In this case, the matter can stay in the metastable state much longer.  As a result, mechanical effects start to dominate over the thermal ones and material fragmentation (cavitation in liquid) occurs under the action of a tensile pressure wave. In other words, the time to mechanical fracture is shorter than the nucleation time ($\tau_{nucl}<\tau_{spal}$). As soon as the time to fracture is exceeded, the material is decomposed into droplets and chunks (separated by vacuum). This mechanical decomposition, described also in Ref.~\onlinecite{Zhigilei, Perez, Lorazo}, concerns the major part of the ablated material in our simulations.  

\begin{figure*}
\includegraphics[width=1.0\textwidth]{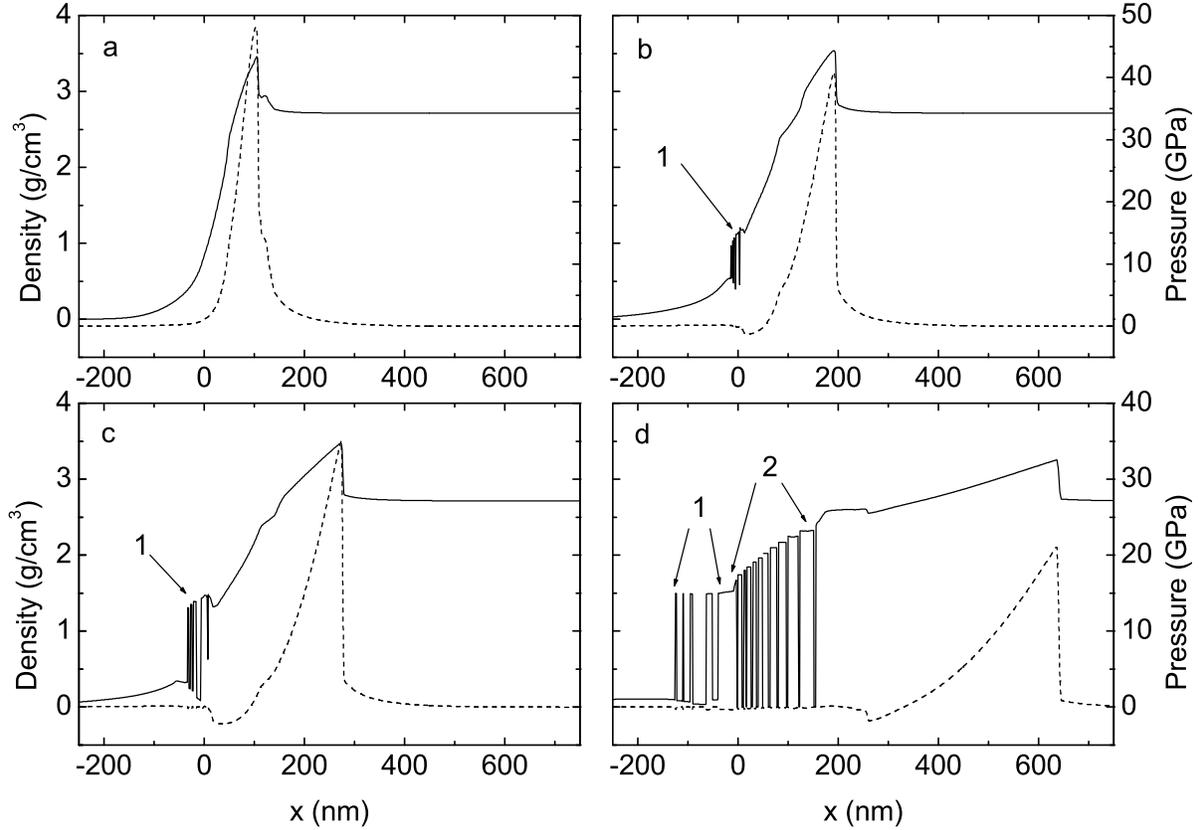}
\caption{\label{fig:2} Contour plots of density (solid) and pressure of heavy particles (dashed) for different time delays after irradiation: (a)---10~ps, (b)---20~ps, (c)---30~ps and (d)---80~ps.  Here, 1---is thermal decomposition zone, 2---is mechanical fragmentation zone.  Laser parameters are the same as in Fig.~\ref{fig:1}.}
\end{figure*}

The dynamics of the target surface layers decomposition is presented in Fig.~\ref{fig:2}.  
Initially, the solid--vacuum interface is located at $x=0$~nm and target occupies $x\geq0$~nm half space.  During the collisions of heated electrons with the lattice its temperature and pressure grows, and thus shock wave forms and propagates into the target (pressure peaks in Fig.~\ref{fig:2}). Pressure release on the free surface of the target results in material expansion in the opposite direction through a rarefaction wave. At a delay $t\sim10$~ps after the beginning of the laser pulse, the density profile is still continuous [Fig.~\ref{fig:2}(a)].  At $t\sim20$~ps, the trajectories originated from the skin-layer reach the binodal curve and penetrate into the metastable liquid region, where thermal decomposition of substance starts [Fig.~\ref{fig:2}(b)].  This process is completed by $t\sim30$~ps [Fig.~\ref{fig:2}(c)]. Then, only mechanical effects, caused by the tensile stress (negative pressure), are involved into the process of decomposition [Fig.~\ref{fig:2}(d)].  By the time $t\sim80$~ps the decomposition of the liquid phase is completed, because the tensile pressure wave intensity drops owing to the energy expenditure on fragmentation (compare minimal pressures in Fig.~\ref{fig:2}(c)--(d) at $x\sim40$ and $x\sim260$~nm, respectively). Completion of the mechanical decomposition corresponds to the final vibrations of the trajectories 5--7 in Fig.~\ref{fig:1} around isobar with zero pressure, similar to recent MD analysis~\cite{Lorazo} and hydrodynamic simulations.~\cite{Colombier06}

The fractions of the target material ablated due to the described mechanisms depend both on the material and on laser parameters. At very small fluences ($F<0.25$~J/cm$^2$ for aluminum) only melting occurs.  When laser fluence is slightly above this value, material fragmentation takes place, in agreement with the previous MD simulations. \cite{Ivanov, Perez}  If we further increase laser fluence, both mechanical and thermal mechanisms play a role.  At larger $F$, all three mechanisms (i) atomization; (ii) thermal decomposition (\textit{critical-point phase separation}); and (iii) mechanical decomposition (\textit{dynamic fragmentation}) occur in different layers of the target.  The third mechanism, however, is found to be dominant in our simulations, accounting about 80\% of ablated mass at all considered laser fluences (from 0.1 to 5~J/cm$^2$).  

\begin{figure}
\includegraphics[width=1.0\columnwidth]{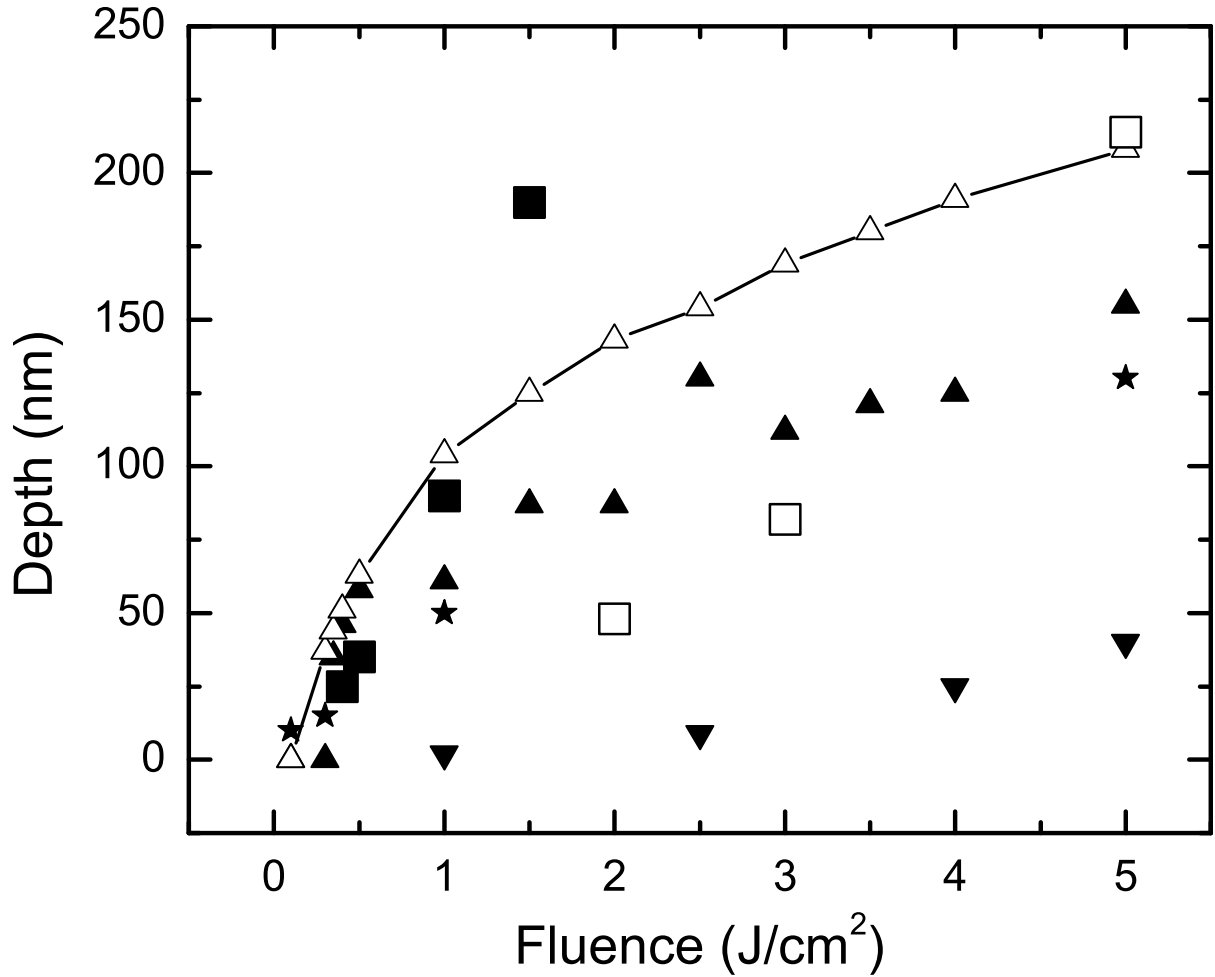}
\caption{\label{fig:3} Present simulation: melted ($\vartriangle$, solid line) and ablated ($\blacktriangle$) depths as a function of laser fluence for aluminum. Experimental data are taken from Ref.~\onlinecite{Amoruso}---($\blacksquare$) and Ref.~\onlinecite{Colombier05}---($\square$). The errors of the experimental data were reported to be of the order of 5--10\%. Results of simulation Ref.~\onlinecite{Komashko}---($\bigstar$) and Ref.~\onlinecite{Vidal}---($\blacktriangledown$) are also shown.}
\end{figure}

Finally, we propose several predictions from our simulations and compare our results with the available experimental findings~\cite{Amoruso,Colombier05} and simulations.~\cite{Komashko,Vidal}  We use a single-shot experimental data for the numerical model verification, though oxidation of aluminum target surface and usage of different methods of measurement result in essential discrepancy in single-shot laboratory findings, see Fig.~\ref{fig:3}. It turns out, that the consideration of only \textit{critical-point phase separation} effect in simulations \cite{Vidal} gives rise to a significant underestimation of the ablation depth (several times) with respect to the experimental values (Fig.~\ref{fig:3}). 
 This underestimation can be explained by the fact that only 10--20\% of the target material is ablated due to these critical effects, whereas most of the ablated material is ejected due to the mechanical fragmentation of the liquid phase.  These results confirm that the decomposed melted zone contributes strongly into the estimation of the ablated depth.
\section{CONCLUSION}
In conclusion, we have presented the results of the hydrodynamic calculations of femtosecond laser ablation with a thermodynamically complete equation of state and with interplay between both kinetic lifetime of metastable liquid state and mechanical time to fracture.  Ablation mechanisms have been investigated as a function of laser fluence.  A correlation between the melted and ablated depths has been revealed. Mechanical decomposition of a metastable liquid layer has been shown to be the dominant ablation mechanism.  In the present work we have considered only aluminum ablation in one dimension.  Similar analysis can be performed for different materials. In addition, an extension of the model to a two-dimensional case with multi-layer materials is underway. 
\section*{ACKNOWLEDGMENTS}
This work was supported by the Russian Foundation for Basic Research (project No. 06-02-17464) and the Council on Grants from the President of the Russian Federation (project No. NSh-3683.2006.2). M.E. Povarnitsyn gratefully acknowledges the financial support from the CNRS, France.


\begin{thebibliography}{9}

\bibitem{R1}
Short Pulse Laser Interactions with Matter: An Introduction, Ed. by P. Gibbon (Imperial College Press, London, 2005). 

\bibitem{R2}
Femtosecond Laser Spectroscopy, Ed. by Peter Hannaford (Springer, 2004).

\bibitem{R3}
Femtosecond Technology for Technical and Medical Applications, Ed. by F. Dausinger, F. Lichtner, H. Lubatschowski (Springer, 2004). 

\bibitem{Nolte}
S. Nolte, B. N. Chichkov, H. Welling, Y. Shani, K. Lieberman, and H. Terkel, Opt. Lett. 24, 914 (1999). 

\bibitem{Amoruso}
S. Amoruso, R. Bruzzese, M. Vitiello, N. N. Nediakov, and P. A. Atanasov, Appl. Phys. 98, 044907 (2005). 

\bibitem{Plech}
A. Plech, V. Kotaidis, M. Lorenc, and J. Boneberg, Nature Physics 2, 44 (2006). 

\bibitem{Colombier05}
J. P. Colombier, P. Combis, F. Bonneau, R. Le Harzic, and E. Audouard, Phys. Rev. B 71, 165406 (2005). 

\bibitem{Colombier06}
J. P. Colombier, P. Combis, A. Rosenfeld, I. V. Hertel, E. Audouard, and R. Stoian, Phys. Rev. B 74, 224106 (2006). 

\bibitem{Rethfeld}
B. Rethfeld, K. Sokolowski-Tinten, D. von der Linde, and S. I. Anisimov, Phys. Rev. B 65, 092103 (2002). 

\bibitem{Anisimov1}
S. I. Anisimov and B. S. Luk'yanchuk, Phys. Usp. 45, 293 (2002). 

\bibitem{Komashko}
A. M. Komashko, M. D. Feit, A. M. Rubenchik, M. D. Perry, and P. S. Banks, Appl. Phys. A 69, S95 (1999). 

\bibitem{Ivanov}
D. S. Ivanov and L. V. Zhigilei, Phys. Rev. B 68, 064114 (2003). 

\bibitem{Anisimov2}
S. I. Anisimov, B. L. Kapeliovich, and T. L. Perel'man, Sov. Phys. JEPT 39, 375 (1974). 

\bibitem{Eidmann}
K. Eidmann, J. Meyer-ter-Vehn, T. Schlegel, and S. H\"uller, Phys. Rev. E 62, 1202 (2000). 

\bibitem{Vidal}
F. Vidal, T. W. Johnston, S. Laville, O. Barth\'elemy, M. Chaker, B. Le Drogoff, J. Margot, and M. Sabsabi, Phys. Rev. Lett. 86, 2573 (2001). 

\bibitem{Glover}
T. E. Glover, J. Opt. Soc. Am. B 20, 125 (2003). 

\bibitem{Zhigilei}
L. V. Zhigilei and B. J. Garrison, J. Appl. Phys. 88, 1281 (2000). 

\bibitem{Garrison}
B. J. Garrison, T. E. Itina, and L. V. Zhigilei, Phys. Rev. E 68, 041501 (2003). 

\bibitem{Schafer}
C. Sch\"afer, H. M. Urbassek, and L. V. Zhigilei, Phys. Rev. B 66, 115404 (2002). 

\bibitem{Perez}
D. Perez, and L. J. Lewis, Phys. Rev. B 67, 184102 (2003). 

\bibitem{Lorazo}
P. Lorazo, L. J. Lewis, and M. Meunier, Phys. Rev. B 73, 134108 (2006). 

\bibitem{Grady}
D. E. Grady, J. Mech. Phys. Solids 36, 353 (1988). 

\bibitem{Kanel}
G. I. Kanel, S.V. Rasorenov, and V.E. Fortov, J. Appl. Mech. Tech. Phys. 25, 701 (1984). 

\bibitem{Miller}
G. H. Miller and E. G. Puckett, J. Comput. Phys. 128, 134 (1996). 

\bibitem{Povarnitsyn}
M. E. Povarnitsyn, T. E. Itina, P. R. Levashov, and K. V. Khishchenko, Appl. Surf. Sci. in press (2007). 

\bibitem{Anisimov3}
S. I. Anisimov and B. Rethfeld, Proc. SPIE Int. Soc. Opt. Eng. 3093, 192 (2002). 


\bibitem{Khishchenko}
K. V. Khishchenko, in Physics of Extreme States of Matter-2005 (IPCP RAS, Chernogolovka, 2005), pp. 170-172. (in Russian) 

\bibitem{LandauLifshits}
L. D. Landau, and E. M. Lifshits, Statistical Physics (Pergamon Press, Oxford, 1980).

\bibitem{Khishchenko1} 	K. V. Khishchenko, S. I. Tkachenko, P. R. Levashov, I. V. Lomonosov, and V. S. Vorob'ev, Int. J. Thermophys. 23, 1359 (2002). 

\bibitem{Oreshkin} 	V. I. Oreshkin, R. B. Baksht, N. A. Ratakhin, A. V. Shishlov, K. V. Khishchenko, P. R. Levashov, and I. I. Beilis, Phys. Plasmas. 11, 4771 (2004).

\bibitem{Frenkel}
J. Frenkel, Kinetic Theory of Liquids (Clarendon Press, Oxford, 1946). 

\bibitem{Skripov}
V. P. Skripov,  Metastable Liquids (New York: Wiley, 1974).

\bibitem{Tkachenko}
S. I. Tkachenko, V. S. Vorob'ev, and S. P. Malyshenko, J. Phys. D: Appl. Phys. 37, 495 (2004).

\bibitem{Surftens}
V. K. Semenchenko, Surface Phenomena in Metals and Alloys, (Pergamon, New York, 1961).

\end{thebibliography}
 \end{document}